\documentstyle[prl,twocolumn,aps]{revtex}
\input epsf
\begin{document}
\draft
\title{Atomic micromotion and geometric forces  in a triaxial magnetic trap}
\author{J. H. M\"uller, O. Morsch, D. Ciampini, M.~Anderlini, R. Mannella,
and E. Arimondo}
\address{INFM, Dipartimento di Fisica, Universit\`{a} di Pisa, Via
Buonarroti 2, I-56127 Pisa, Italy}
\date{\today}
\maketitle
\begin{abstract}
Non-adiabatic motion of Bose-Einstein condensates of
rubidium atoms arising from the dynamical nature of a time-orbiting-potential
(TOP) trap was observed experimentally. The orbital micromotion of the
condensate
in velocity space at the frequency of the rotating bias
field of the TOP was detected by a time-of-flight method. A dependence of the
equilibrium position of the atoms on the sense of rotation of the bias
field was observed. We have compared our experimental findings with
numerical simulations.
The nonadiabatic following of the atomic spin in the trap rotating
magnetic field produces geometric forces acting on the trapped atoms.
\end{abstract}

\narrowtext Magnetic trapping has become a standard technique for
confining and evaporatively cooling ultra-cold samples of neutral
atoms. Up to now, BEC in alkali atoms has been achieved in two
different kinds of magnetic traps: static traps, and dynamic ones,
also called time-orbiting-potential (TOP) traps\cite{inguscio}. In
the latter, introduced in \cite{toptrap}, a quadrupole magnetic
field is modified by a rotating homogeneous bias field which
causes the quadrupole field to orbit around the origin, thus
eliminating the zero of the magnetic field at the trap center. In
the original scheme with cylindrical symmetry~\cite{toptrap}, the
bias field rotates in the symmetry plane of the quadrupole field,
whereas in a later version~\cite{hagley} the plane of rotation
contains the quadrupole symmetry axis, which yields a completely
asymmetric (triaxial) magnetic trap. In the present work, precise
measurements on the motion of a $^{87}$Rb condensate in the triaxial
TOP trap have allowed us to discover new dynamical features,  beyond
the approximations
previously applied in the theoretical treatment of
time-dependent magnetic traps.

In the adiabatic approximation the spins of the magnetically
trapped atoms follow the instantaneous direction of the magnetic
field. Generally speaking, an adiabatic approximation is applied
when the evolution of one set of variables of a system is fast
compared to the time scale of a second one. While the full
dynamics can be quite complicated and hard to analyze, in a useful
approximation the fast motion is solved for fixed values of the
slow coordinates, and then the slow dynamics is considered to be
governed by the {\em average} values of the fast motion. In a
classical description, the energy of the fast motion calculated
for fixed values of the variables of the slow system acts as a
potential in which the evolution of the slow system takes place.
In an improved approximation, reaction forces act on the slow
dynamics. They are related to the geometric phase and to the
geometric gauge fields ~\cite{berry}. In the classical case, the
physical origin of the geometric forces has been studied, e.g.,
for atomic systems with spins~\cite{aharonov}. For atomic motion
in an inhomogeneous magnetic field, the slow atomic external
degree of freedom acts as a ``guiding center'' for the evolution of
the fast precessing spin variables coupled to the magnetic field.
The resulting geometric forces are usually very small and hard to
detect. Analogous terms in the calculation of molecular spectra
beyond the Born-Oppenheimer approximation do, however, lead to
measurable shifts of energy levels, underlining the usefulness of
the concept~\cite{berry}.

In the TOP trap, the guiding-center atomic motion within the
time-dependent inhomogeneous magnetic field acquires an additional
dynamic feature associated with the periodic evolution of the
rotating field, feature not present in the analyses
of~\cite{aharonov}. Because the frequency of the bias field
rotation is typically much larger than the vibrational frequencies
of the atoms in the trap, in an additional approximation a
time-average over the period of the rotating field is usually
performed. Application of both adiabatic and time-average
approximations predicts a description of the atomic motion in
terms of a time-independent potential. A number of theoretical
studies have considered the role of the nonadiabatic
dynamics~\cite{yukalov} or of the time-dependent potential on the
atomic evolution in a TOP trap~\cite{minogin}. Experimental
investigations, however, have restricted their attention mainly to
the atomic response to the time-independent adiabatic potential.

The breakdown of the time-average approximation
and of the adiabatic spin following approximation, and
the evidence for additional geometric forces are visible in the
following measurements on a
$^{87}$Rb condensate~\cite{footnote2} in a triaxial trap:
 i) the atomic micromotion at the
frequency of the bias field, ii) the change of the atomic
equilibrium position in the trap when inverting the sense of
rotation of the bias field\cite{hall1}, iii) the dependence of the
rotation-sensitive equilibrium position on the applied magnetic
fields. The dependence of the equilibrium position on the sense of
rotation of the bias field does not violate any fundamental
symmetry law, as the handedness of the Larmor precession  for an
atom in a trapped state defines the sense of rotation in
the atomic evolution.
In our experimental investigation,as for other
TOP traps, the typical
conditions for applying the adiabatic and time-averaged
approximations are not violated~\cite{othertoptraps}. A precise
detection of the micromotion, however, allowed us to perform
measurements of the non-adiabatic effects present in the TOP trap.
We have compared our results with analytical models and numerical
simulations based on a completely classical description, obtaining
good agreement.

\begin{figure}\centering\begin{center}\mbox{\epsfxsize 2.8 in
\epsfbox{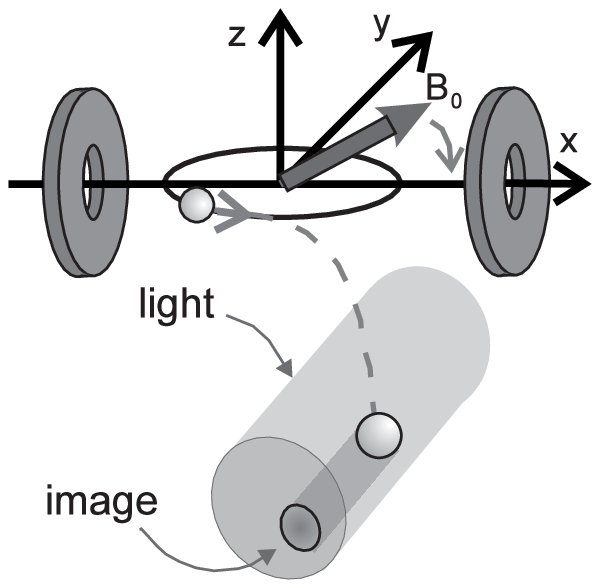}}
\end{center}
\caption{Schematic of the static and rotating magnetic fields
confining the condensate, and of the shadow-imaging method.
Imaging along the $y$-axis allowed us to measure the velocity of
the atoms along the $x$ and $z$ axes. For our trap parameters the
spatial amplitude of the micromotion in the trap was less than
$1\,\mathrm{\mu m}$, well below our resolution limit of
$5\,\mathrm{\mu m}$.} \label{setup}
\end{figure}

Our experimental apparatus~\cite{jphysbpaper} is based on a
double-MOT system with two cells connected by a glass tube to
capture, cool and transfer atoms into a magnetic TOP trap. The
latter consists of a pair of quadrupole coils oriented along a
horizontal axis and two pairs of bias-field coils, one
incorporated into the quadrupole coils and the other along a
horizontal axis perpendicular to that of the quadrupole coils, as
shown in Fig.~\ref{setup}. After transferring the atoms into the
TOP-trap, we evaporatively cool them first by reducing the radius
of the quadrupole orbit and finally by applying an RF-field. Once
Bose-Einstein condensation has occurred in the $|F=2,
m_{F}=2\rangle$ state, we adiabatically changed the quadrupole
gradient and the bias field to the desired values. The
instantaneous magnetic field seen by the atoms is:
\begin{eqnarray}\label{magfield}
\bf{B} &=& \bigl[2b^{\prime}x+B_0 \cos\omega_{0}t\bigr]{\bf \hat{i}}+
\bigl[-b'y+B_0
\sin\omega_{0}t\bigr] {\bf \hat{j}}-b^{\prime}z{\bf \hat{k}}
\end{eqnarray}
where $b'$ is the quadrupole gradient along the $z$-axis, and
$B_0$ and $\omega_{0}$ are the magnitude and the frequency of
the bias field, respectively.
 Application of the adiabatic and time-average
approximations leads to an oscillatory motion of the atoms in a
time-independent potential with a harmonic frequencies
$\omega_x:\omega_y:\omega_z$ of $2:1:\sqrt{2}$, where
$\omega_{x}=b^{\prime}\sqrt{2\mu/mB_{0}}$, $m$ being the mass and
$\mu$ the
magnetic moment of an atom in the trapped state\cite{hagley}.
Lifting the time-average approximation yields
the atomic micromotion, because  the atoms experience a
time-dependent potential at frequency $\omega_{0}$ and its
harmonics. Expansion of the potential energy $U$ for atoms close
to the trap center gives additional linear terms $\Delta U$
modulated at the frequency $\omega_{0}$ on top of the effective
harmonic potential, with
\begin{equation}
\Delta U = \mu b^{\prime}\left(2x \cos\omega_{0}t-y \sin
\omega_{0}t\right).
\end{equation}
\begin{figure}\begin{center}\mbox{\epsfxsize 3.5in \epsfbox{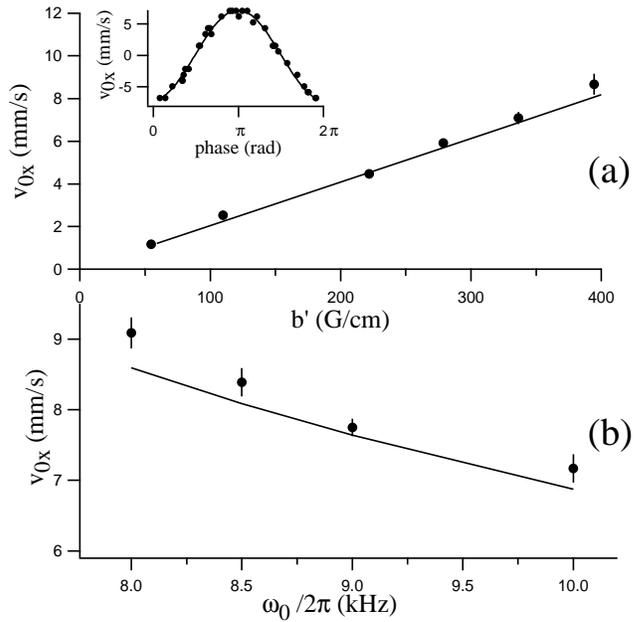}}
\end{center}
\caption{Dependence of the micromotion velocity $v_{0x}$ on  $b'$
at $\omega_{0}=2\pi\times10^{4}$ s$^{-1}$ in (a) and  on
$\omega_{0}$ in (b). In (b), $b'$ =$329\,\mathrm{Gcm^{-1}}$ and
$B_{0}$ =28 G. In (a), for gradients larger than
$200\,\mathrm{Gcm^{-1}}$, $B_0=28\,\mathrm{G}$ was used, whereas
for smaller gradients $B_0$ was reduced in order to maintain a
small gravitational sag. Continuous lines from
Eqs. (\ref{velocities}). The inset in (a) shows the velocity
dependence on the phase of the bias field at which the trap was
switched off (zero phase corresponds to the bias field pointing
{\it along} the direction of observation).} \label{micromotion}
\end{figure}
For trap oscillation frequencies well below  $\omega_{0}$,
separation of the time scales allows us to write down the atomic
micromotion  at $\omega_{0}$ as a small shaking motion on top of
the secular motion in the trap. Balancing the restoring force from
the $\Delta U$ potential with the centrifugal force of the
resulting orbital motion yields a periodic micromotion with
velocity amplitudes
\begin{equation}
v_{0x}=\frac{2\mu b^{\prime}}{m\omega_{0}}\quad {\mathrm and} \quad
v_{0y}=\frac{\mu
b^{\prime}}{m\omega_{0}}. \label{velocities}
\end{equation}
In order to observe the micromotion of the atoms at the
TOP-frequency, the trap fields were switched off in a time (less
than $20\,\mathrm{\mu s}$)  short compared to the period
of the TOP-field \cite{footnote1}. The instant at which the
trapping fields were switched off was determined to better than
$1\%$ of a TOP-cycle. After switching off the quadrupole field at
a well-defined phase of the rotating bias field and after a variable
delay of up to $8\,\mathrm{ms}$, we imaged the absorption shadow
of the falling cloud onto a CCD-chip. Because of the small spatial
amplitude of the micromotion, the measured position of the atoms
along the $x$-axis after the time-of-flight was determined solely
by their initial velocity $v_{0x}$ in that direction (see
Fig.~\ref{setup}).

A typical plot of $v_{0x}$ as a function of the bias-field phase
is shown in the inset of Fig.~\ref{micromotion}. A velocity
amplitude of $5\,\mathrm{mm\,s^{-1}}$ corresponds to  a spatial
amplitude of the micromotion around $100\,\mathrm{nm}$.  Also
shown in Fig.~\ref{micromotion} is the dependence of $v_{0x}$ on
the quadrupole gradient and on the frequency of the rotating bias
field. We found no dependence of $v_{0x}$ on $B_0$. All of these
dependencies are in agreement with
Eq.~(\ref{velocities})(continuous lines in
Fig.~\ref{micromotion}). We checked by numerical integration of
the equations of motion in the adiabatic approximation that the
finite switching time of the trapping field leads to a reduction
of the measured micromotion velocity of at most a few percent compared to
Eq.~\ref{velocities}.

\begin{figure}\begin{center}\mbox{\epsfxsize 3.5in \epsfbox{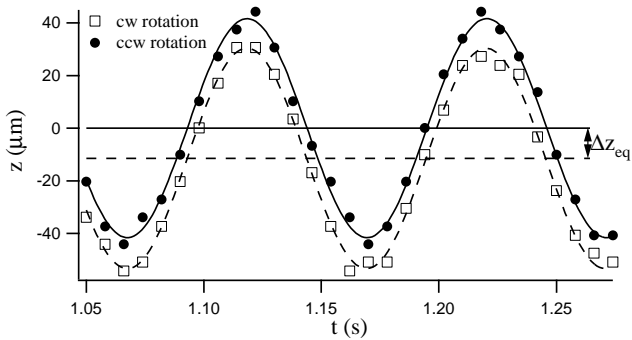}}
\end{center}
\caption{Centre-of-mass oscillations of atoms in the TOP-trap for clockwise
and counterclockwise
rotation of the bias field. The gradient $b'$  and the bias field $B_0$
were 20.8 Gcm$^{-1}$ and 4
G, respectively. The equilibrium positions for the two cases are indicated
by the horizontal
lines.}\label{oscillations}
\end{figure}

When the sense of rotation of the bias field was reversed, the
phase of the measured velocity was shifted by $180$ degrees, in
accordance with the physical picture of the atoms being `dragged'
along by the instantaneous position of the quadrupole center. On
top of this phase shift, we observed a shift in the vertical
equilibrium position of the atoms in the TOP-trap.
 We achieved a high resolution of the equilibrium position by using very
 small condensate clouds  and by exciting dipole oscillations of the cloud
along the $z$-axis.
 Typical results of such measurements for the two senses of rotation of the
bias field are shown in
Fig.~\ref{oscillations}. Fitting a
sinusoidal curve to the position data we determined the absolute centre of
the oscillations to
within $5\,\mathrm{\mu m}$ and the relative position of the centers to
within $3\,\mathrm{\mu m}$.
 Identical conditions for both senses of rotation were ensured by
powering the bias coils from two phase-locked signal generators
and shifting their relative phase by $180$ degrees to reverse the
sense of rotation whilst keeping all other parameters constant.

\begin{figure}\begin{center}\mbox{\epsfxsize 3.5in \epsfbox{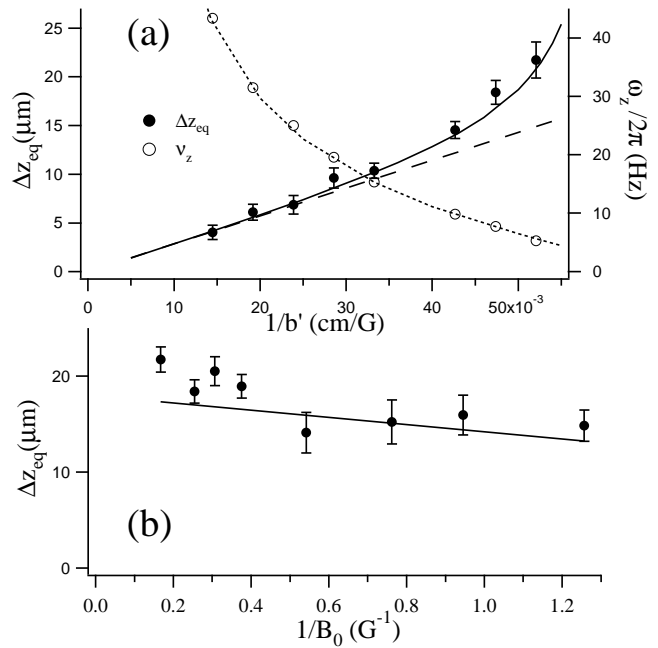}}
\end{center}
\caption{(a) shows the dependence on the quadrupole gradient $b'$ of the
splitting $\Delta z_{\rm eq}$ of the equilibrium positions (filled
symbols) for opposite senses of rotation of the bias field, and
dependence of
the oscillation frequencies (open symbols). The solid and dotted
lines are the results of a numerical simulation {\it without} the
adiabatic approximation. The dashed line is the prediction of
Eq.\ref{split}; $B_0=4\,\mathrm{G}$ for all points. In (b) we plot the
dependence
of the splitting (symbols: data, solid line: numerical simulation)
on the inverse of the bias field strength for fixed gradient
($b'=20\,\mathrm{Gcm^{-1}}$).} \label{splitting}
\end{figure}

The measured dependence of the spatial separation $\Delta z_{\rm eq}$ of
the two equilibrium positions on the inverse of the quadrupole
gradient, $1/b'$, is shown in Fig.~\ref{splitting} (a). For large
gradients, we find a linear dependence of $\Delta z_{\rm eq}$ on $1/b'$,
whereas for the smallest gradients the splitting data deviate
significantly from a straight line as $b'$ approaches the minimum
gradient, $b'_{min}=15.3\,\mathrm{Gcm^{-1}}$, necessary to balance
gravity. In Fig.~\ref{splitting} (b) the measured dependence of
$\Delta z_{\rm eq}$ on the inverse of the bias-field $B_{0}$ is shown. Also
plotted in Fig.~\ref{splitting} (a) is the dependence of
$\omega_{z}$ on $1/b'$, allowing a calibration of $b'$ with an
accuracy of two percent over a very large range.

We have  solved numerically the coupled differential equations of
\cite{gov} for the center of mass motion and atomic spin within
our TOP trap field. The numerical simulations provided an
excellent fit for the results of Fig.
\ref{splitting}\cite{noteell}. To find the equilibrium position, a
transient damping was introduced in the numerical integration of
the evolutions  for the center-of-mass and/or the spin. We have
verified that damping rates corresponding to the  trap  lifetime
do not modify the atomic micromotion and the non-adiabatic
effects. However the role of additional damping mechanisms, for
instance the viscous heating of~\cite{toptrap}, should be tested.

To interpret the dependence of the  splitting $\Delta z_{\rm eq}$
on the field gradient, a transformation to a frame rotating
synchronously with the bias field may be applied, as
in~\cite{hall1,gov}. For the special case of a cylindrical TOP
trap this transformation completely removes the time dependence.
The frame rotation acting on the spin variables introduces a
fictitious (geometric) magnetic field  ${\bf B}_{g}=-m_{\rm F} \hbar
\omega_{0}\hat{\bf k}/\mu$ along the  rotation axis, with $m_{\rm
F} \hbar$ the atomic internal angular momentum. In a refined
adiabatic approximation the spin is aligned with the effective
field, i.e. the sum of the real and geometric magnetic fields. The
occurrence of transverse components of the the spin with respect
to the real magnetic field lies at the heart of the concept of
geometric forces. The geometric field  shifts vertically the
plane of the quadrupole orbit. The difference $\Delta z_{\rm eq}$ in the
equilibrium positions for cw and ccw rotations of the bias field
 is
\begin{equation}
\label{split}
 \Delta z_{\rm eq} =2\frac{m_{\rm F}\hbar |\omega_{0}|}{\mu b^{\prime}}.
\end{equation}
Gravity shifts the equilibrium position $z_{\rm  eq}$ downwards,
but for a cylindrical trap the above result for $\Delta z_{\rm
eq}$ is not modified. The expressions for the geometric force
of~\cite{aharonov} give rise to an equivalent result for the
geometric field and the splitting $\Delta z_{\rm eq}$. While for
the cylindrical TOP trap the transformation to the rotating  frame
takes care completely of the spin dynamics, for the triaxial TOP
trap the rotating frame transformation does not result in a
time-independent problem. In fact, at low $b'$ values, approaching
$b'_{\rm min}$, the splitting data of Fig. \ref{splitting}(a)
 deviate from the behavior described by Eq. (\ref{split}). Higher order
geometric
 terms derived in~\cite{aharonov} lead to corrections  several orders of
magnitude too small to explain these deviations. Numerically we
found that at zero gravity also for the triaxial TOP trap $\Delta
z_{\rm eq}$ is well described by Eq. (\ref{split}). This leads us
to conjecture that gravity modifies the fast correlation between
spin orientation and center-of-mass motion responsible for the
geometric forces.

In summary, in a TOP trap the micromotion and  the splitting of
the equilibrium positions of the atoms for oppositely rotating
bias fields are accounted for by a lifting of the time-average and
adiabatic approximations. The  dependence of the splitting
 on the magnetic field gradient arises from the
nonadiabatic
 characteristics of the
TOP trap. The micromotion in a TOP-trap is of relevance for
experiments where an accurate control of the condensate residual
motion on the recoil velocity scale ($5\,\mathrm{mm\,s^{-1}}$) is
required, e.g. in atom lasers or condensate transfer.  The
observed dependence of the equilibrium position on the bias field
rotation hopefully motivates theoretical efforts to calculate
higher order corrections to  the adiabatic spin evolution in the
TOP-trap configuration. Finally it should be tested, theoretically
and experimentally, whether the condensate response to the
parallel transport  of the atomic spin in a TOP trap,
which gives rise to the geometric forces, may influence also the internal
dynamics of the condensate.

We thank G. Smirne for help in the data acquisition.  O.M. gratefully
acknowledges financial
support from the European Union (TMR Contract-Nr. ERBFMRXCT960002). This
work was supported by the
INFM ``Progetto di Ricerca Avanzata'', and by the CNR ``Progetto Integrato''.


\begin{references}
\bibitem{inguscio} Recent reviews  by W. Ketterle,
{\it et al.}, and by E.~Cornell, {\it et al.} in {\it
Bose-Einstein condensation in atomic gases"}, edited by M. Inguscio,
S.~Stringari and C.~Wieman
(IOS Press, Amsterdam) 1999.
\bibitem{toptrap} W.~Petrich,{\it et al.}, Phys. Rev. Lett. {\bf 74,} 3353
(1995).
\bibitem{hagley}
E.W.~Hagley,{\it et al.}, Science {\bf 283}, 1706 (1999).
\bibitem{berry} M.V. Berry, in {\it Geometric Phases in
Physics}, edited by A.~Shapere and F.~Wilczek (World Scientific, Singapore,
1989) pp. 1-28.
\bibitem{aharonov} Y.~Aharonov and A.~Stern, Phys. Rev. Lett. {\bf
69}, 3593 (1992); R.G.~Littlejohn and S.~Weigert, Phys. Rev. A {\bf
48}, 924 (1993); M.V.~Berry and J.M.~Robbins, Proc. Roy. Soc. London Ser.
A {\bf 442}, 641 (1993);  M.V.~Berry, {\it ibid.}  {\bf 452}, 1207 (1996).
\bibitem{yukalov} V.I. Yukalov, Phys. Rev. A {\bf 56}, 5004 (1997);
V.E.~Shapiro, {\it ibid.} {\bf 60}, 719 (1999).
\bibitem{minogin} A.B.~Kuklov, {\it et al.},  Phys. Rev. A{\bf 55}, 488
(1997); V.G.~Minogin, {\it et al.}, {\it ibid.} {\bf 58}, 3138 (1998).

\bibitem{footnote2} We used condensates mainly because their small size
greatly facilitated precise
position measurements. Experiments with evaporatively cooled thermal clouds
yielded the same results within the
experimental uncertainty.
\bibitem{hall1} The rotation sensitive condensate position was
  discussed by D.S.~Hall, {\it et al.}, Proc. SPIE {\bf 3270}, 98 (1998).
It is also contained implicitly in in the analytical model of \cite{gov}.
 \bibitem{gov} S. Gov and S.~Shtrikman, J. Appl. Phys. {\bf 86}, 2250
(1999).
\bibitem{othertoptraps}TOP traps have been
used for BEC in \cite{toptrap,hagley} and by D.J.Han, {\it et al.},  Phys.
Rev. A {\bf 57} R4114
(1998); J.L.~Martin, {\it et al.}, J. Phys. B: Atom. Mol. Opt. Phys. {\bf
32}, 3065 (1999);
J.~Arlt, {\it et al.}, {\it ibid.} {\bf 32} 5861 (1999).
\bibitem{jphysbpaper} J.H.~M\"uller,{\it et al.}
J. Phys. B: Atom. Mol. Opt. Phys. in press and condmat/0005133.
\bibitem{footnote1}The switch-off time of the bias field, limited by the
quality factor of tank
circuits was around
$150\,\mathrm{\mu s}$. However, fast switching of the bias field was
not required as the atoms are no longer trapped once the
quadrupole field is extinguished.
\bibitem{noteell}We should mention that the agreement between
simulations and experiments becomes less satisfactory for very
 elliptical bias field
in the $x-y$ plane.
\end{references}
\end{document}